\documentclass[a4paper,aps,prb,twocolumn,floatfix,showpacs,superscriptaddress,footnoteinbib]{revtex4-1}
\usepackage{graphics}
\usepackage{epsfig}
\usepackage{bbm}
\usepackage{times}
\usepackage{xcolor}   
\usepackage{amsfonts}
\usepackage{amsmath}
\DeclareMathOperator{\sech}{sech}
\usepackage{amssymb}
\usepackage{amsthm}
\usepackage{hyperref} 
\usepackage{wasysym}
\usepackage{bm} 

\newcommand{\sgn}{\mathop{\mathrm{sgn}}}

\begin{document}

\title{
Volkov-Pankratov states on the edge of a quantum spin Hall system
}

\author{Vivekananda Adak}
\affiliation{Department of Physical Sciences, IISER Kolkata, Mohanpur, West Bengal 741246, India.}
\author{Subhadeep Chakraborty}
\affiliation{Department of Physical Sciences, IISER Kolkata, Mohanpur, West Bengal 741246, India.}
\author{Krishanu Roychowdhury}
\affiliation{Department of Physics, Stockholm University, SE-106 91 Stockholm, Sweden.}
\affiliation{Max-Planck-Institut f\"{u}r Physik komplexer Systeme,
N\"{o}thnitzer Strasse 38, 01187 Dresden, Germany.}
\author{Sourin Das}
\affiliation{Department of Physical Sciences, IISER Kolkata, Mohanpur, West Bengal 741246, India.}
\email{vivekanandaaadak@gmail.com, sc20rs001@iiserkol.ac.in, krishanu.1987@gmail.com, sourin@iiserkol.ac.in}

\begin{abstract}
Volkov-Pankratov (VP) states are a family of sub-gap states which appear at the smooth interface/domain wall between topologically distinct gapped states. We study the emergence of such states in the edge spectrum of a quantum spin Hall system subjected to a smoothly varying mass term (Zeeman field) that switches sign at a given spatial point. Both the VP states at non-zero energy and the zero energy Jackiw-Rebbi mode stay localized at the interfacial region, however, the former feature several distinctive signatures compared to the latter such as non-trivial spin textures that can be characterized by a winding number in real space. On applying an electric field, the texture deforms leaving its winding number unaltered. Moreover, the VP states exhibit an intriguing interplay between the electric and the magnetic field with a collapse of the spectrum onto the zero mode when they are equal in magnitude. Quantum transport simulations on a 2D lattice model are performed to undergird our theoretical prediction. 
\end{abstract}
\maketitle

\section{Introduction}

Topological quantum matter exhibits a myriad of unconventional phenomena -- perhaps the most celebrated one is the {\it bulk-boundary correspondence} \cite{shen2012topological} \cite{essin2011bulk} when robust boundary modes emerge due to a topologically nontrivial bulk. Burgeoning experimental activities over the last decades have illuminated possibilities of leveraging this effect in device applications as well  \cite{dolcini2011full, zhang2003spin, roth2009nonlocal, krueckl2011switching, sinova2015spin}. A gapped bulk can harbor a nontrivial topology when the underlying Hamiltonian admits certain spin-orbit interactions (leading to quantum spin Hall insulators \cite{kane2005quantum, kane2005z, zhang2003spin, roth2009nonlocal, sinova2015spin, vzutic2004spintronics}, topological insulators\cite{hasan2010colloquium, moore2007topological} ) or breaks time-reversal invariance (such as the case of a Chern insulator \cite{regnault2011fractional}). By tuning appropriate parameters of the Hamiltonian, one can invert the sign of the bulk gap $\Delta$ near the Fermi energy, thereby creating a domain-wall type configuration in $\Delta$ at a heterojunction setup comprising a junction between a topologically nontrivial bulk and a trivial one. Detecting these modes in experiments renders a direct avenue to probe the topology of the bulk. 

For the specific case of a 1D system, an abrupt change (of the sign) in the profile of the gap $\Delta({\bf x})$ [{\it e.g.}, when $\Delta({\bf x})\propto {\rm sgn}({\bf x})$] would lead to the formation of bound states that are exponentially localized at the interface. A prominent example would be the Jackiw-Rebbi zero mode  \cite{shen2012topological, jackiw1976solitons, rajaraman1982solitons, su1979solitons, su1980soliton} found either in the Su-Schrieffer-Heeger chain or at the magnetic domain walls in the helical edge states of a quantum spin Hall state (QSHS). In distinction, when $\Delta({\bf x})$ varies smoothly, sub-gap states can appear also at non-zero energies whose count depends on the width of the domain wall, identified as the Volkov-Pankratov (VP) states \cite{pankratov1987supersymmetry, pankratov1991energy}. Volkov and Pankratov noted these states while studying interfacial phenomena in ferroelectric compounds in presence of a magnetic field. They further pointed out a supersymmetric structure \cite{pankratov1987supersymmetry} of the underlying effective Hamiltonian for a generic band inversion problem allowing for a prediction of such states without requiring any specific form of the potential at the transition region. Following that trail, numerous topological heterojunction setups have been theoretically studied involving graphene nanoribbons, topological superconductors, Weyl semimetals, and recently a topological-trivial semiconductor heterojunction has been experimentally explored \cite{mahler2019interplay, alspaugh2020volkov, mukherjee2019dynamical, van2020volkov, inhofer2017observation}. 

The central topic of this article is to investigate the properties of the VP states on the edges of a quantum spin Hall insulator. A simple model to describe such a system is the Bernevig-Hughes-Zhang (BHZ) model \cite{bernevig2006quantum} of HgTe quantum well \cite{konig2007quantum} where adjusting merely the well width results in a band inversion, driving the system into a topological insulator state \cite{inhofer2017observation, lu2020dirac, mahler2021massive}. This is a time-reversal symmetric system with the edge modes having a conserved spin quantum number $S_z$ locked with their momentum, viz., if $\uparrow$ spins ($S_z = +1$) flow along $+k$, $\downarrow$ spins ($S_z = -1$) would flow along $-k$. These are known as the helical edge states (HES) having linear dispersions around the $\Gamma$ point \cite{adak2022spin, chen2016pi, rzeszotarski2019electron, gresta2019optimal, liu2011helical, wu2006helical}. The HES of the QSHS are described by massless Dirac fermions where the Fermi sea carries a persistent spin current. Exposing them to a spatially varying transverse magnetic field opens up a gap via Zeeman coupling proportional to the strength of the magnetic field and gives way to surface-localized VP states. The article focuses to explore these boundary modes for which it is sufficient to consider the effective surface Hamiltonian that solely describes the HES of the QSHS. A key finding is establishing the existence of nontrivial spin texture of the VP states in real space with helicoidal winding along the edge, characterized by winding numbers that are intimately connected to the energy quantum numbers. We demonstrate that an in-plane electric field provides an interesting handle for the positions of the winding centers along the edge. Notably, we find that the energy gap between successive VP states gradually shrinks as the strength of the electric field increases. At a critical value of the electric field which is determined by the gradient of the transverse magnetic field, a full collapse of the entire sub-gap spectrum gives rise to an exponentially degenerate manifold of zero modes within the gap.

The rest of the paper is organized as follows. In Section~\ref{sectwo}, we will present model calculations to demonstrate the appearance of the VP states in the edge spectrum of a QSHS setup in presence of a linearly varying magnetic field $B(x)$ as well as a hyperbolic profile of $B(x)$. Alluding to the supersymmetric structure of the problem we will further discuss the effects of the inclusion of an electric field. We will show how it can be used to manipulate the spectrum of the bound states and their spin texture in real space. Section~\ref{secthree} contains the results from transport simulations that are performed using the KWANT package on a 2D lattice model which hosts the QSHS. Finally, we conclude in Section~\ref{secfour}.

\section{QSHS subjected to smooth potentials: emergence of the VP states}\label{sectwo} 

The following model is to demonstrate the emergence of VP states for a smooth domain wall potential separating two phases related by a band inversion. As a specific setup, let us consider a QSHS where the HES are subjected to a smoothly varying magnetic field $B(x)$ opening a spatially varying mass gap. The two phases here are characterized by the sign of this mass gap which can be inverted by changing the sign of the magnetic field. We construct a domain wall configuration of $B(x)$ around $x=0$ such that $B(-x)=-B(x)$ separating the two phases at $x<0$ and $x>0$. The spins of the free HES are polarized along the $z$-direction which is perpendicular to the plane hosting the QSHS while the magnetic field is taken in-plane and oriented along the $x$-axis.

The Hamiltonian governing the dynamics of the free HES (extended from $x = -\infty$ to $x = +\infty$, $x$ representing an intrinsic one-dimensional coordinate along the edge) is a Dirac Hamiltonian 
\begin{align}\label{hamsimple1}
 {\cal H}_{\rm QSH} = \int {\rm d}x~ \Psi^\dagger H \Psi~;~ H = -i\hbar v_F \sigma_z \partial_x,
\end{align}
$v_F$ being the Fermi velocity of the electrons on the edge and $\Psi \equiv \begin{pmatrix} \psi_R & \psi_L \end{pmatrix}^T$ denotes the annihilation operator for the right ($R$) and the left ($L$) moving electrons ($\sigma_i$ are the Pauli matrices). The in-plane magnetic field facilitating backscattering between the two helical edges, in this basis, takes the form 
\begin{align}\label{hamsimple2}
 {\cal H}_{\rm B} = \int {\rm d}x~ \Psi^\dagger H_B(x) \Psi~;~ H_B(x) = g\mu_B B(x)\sigma_x,
\end{align}
where $g$ is the $g$-factor of the electron, and $\mu_B$ is the Bohr magneton. For the rest of the article, we will use the notion of a magnetic potential $M(x)=g\mu_B B(x)$ to simplify notations. Moreover, we will be interested to explore how the finite-energy sub-gap states respond to an electric field that can either be applied externally or can arise naturally in the system from spontaneous defragmentation of the QSHS into regions of distinct spin-orbit fields respecting time-reversal symmetry but breaking $S_z$ conservation \cite{adak2020spin}. The result of the electric field $E(x)$ can be incorporated via including a space-dependent chemical potential $\mu(x)$ (such that $E=-\partial_x \mu$) that leads to a total Hamiltonian of the form
\begin{align}\label{hamsimple3}
 {\cal H} = \int {\rm d}x~ \Psi^\dagger \left[-i\hbar v_F \sigma_z \partial_x + M(x)\sigma_x + \mu(x) \right] \Psi.
\end{align}
Note here both $M(x)$ and $\mu(x)$ are smooth functions of $x$, and to obtain analytic solutions of the eigenvalue problem ${\cal H}\psi=E\psi$, we will restrict to $\mu(x)\propto M(x)$ when an electric field is applied.

In what follows, we will discuss the effect of bounded and unbounded magnetic potential $M(x)$, both in the absence and presence of an electric field. The strength of the electric field is such adjusted that $\mu$ always stays within the gap set by the magnetic potential $M(x)$. As we will see, the VP states display an interesting response to the electric field in distinction to the Jackiw-Rebbi mode, namely, the VP modes shift in the real space along the direction of the applied field in a highly nonlinear fashion.  

\subsection{SUSY quantum mechanics}\label{sectwoA}

In absence of an electric field, the Dirac problem at hand has a supersymmetric structure \cite{pankratov1987supersymmetry}. The Hamiltonian 
\begin{align}\label{hamsimple4}
 {\cal H} = \int {\rm d}x~ \Psi^\dagger \left[-i\hbar v_F \sigma_z \partial_x + M(x)\sigma_x \right] \Psi
\end{align}
can be brought to a chiral form via a unitary transformation
\begin{align}\label{hamsimple5}
 {\cal H} = \int {\rm d}x~ \tilde{\Psi}^\dagger \begin{bmatrix}
   & A^\dagger \\
   A & 
 \end{bmatrix}
 \tilde{\Psi}~;~ A^\dagger = -\hbar v_F \partial_x + M(x).
\end{align}
Denoting the chiral matrix in Eq.~\ref{hamsimple5} as $\tilde H$, the bound state solutions of ${\tilde H}\psi=E\psi$, that are of the form $\psi=\begin{bmatrix}\theta & \phi \end{bmatrix}^T$, are referred to as the VP states. We can decouple the equations for $\theta$ and $\phi$ by squaring $\tilde H$ which produces two Schr\"{o}dinger equations 
\begin{align}\label{hamsimple6}
\left[\partial_x^2-(M/\hbar v_F)^2 + \partial_x M/\hbar v_F - E^2 \right]\theta &= 0, \nonumber \\
\left[\partial_x^2-(M/\hbar v_F)^2 - \partial_x M/\hbar v_F - E^2 \right]\phi &= 0.
\end{align}
In the language of supersymmetric quantum mechanics, the function $M/\hbar v_F$ is known as the superpotential whereas the potentials $(M/\hbar v_F)^2 \pm \partial_x M/\hbar v_F$ are the supersymmetric partner potentials \cite{cooper1995supersymmetry}. The associated Schr\"{o}dinger Hamiltonians admit only non-negative energy eigenvalues (which we index with an integer $N$) and at least one of them must have zero modes (signature of an unbroken supersymmetry \cite{cooper1983aspects}). 

We will now discuss two cases (i) an unbounded superpotential where $M^2(x)\propto x^2$ -- a harmonic oscillator potential, and (ii) a bounded superpotential where $M^2(x)\propto \tanh^2(x)$ -- referred to as the Rosen-Morse potential \cite{gangopadhyaya2017supersymmetric, dabrowska1988explicit}, and note distinct signatures in the resulting VP states. We will then apply an electric field in both setups to see how the energy spectrum and spatial texture of these states are influenced with identifiable signatures in transport measurements. Note these are some of the well-known examples of a reflection-less potential where quantum transport studies of the scattering states may also reveal interesting effects.  

\subsection{Unbounded $M(x)$, $\mu(x)=0$}\label{sectwoB}

Let us first consider a simple unbounded superpotential $M(x)=\alpha x$ to illustrate the physics associated with the emergence of VP states when no electric field is present {\it i.e.}, $\mu(x)=0$. Such a magnetic potential arises from an odd-parity magnetic field linearized around a given spatial point $x=0$ and extended throughout the HES (from $x=-\infty$ to $x=+\infty$), namely $B(x)=B_0 x$. 
The operator $A^\dagger$ in this case becomes proportional to the conventional bosonic ladder operator $A^\dagger=\sqrt{2\alpha \hbar v_F}~ a^\dagger$ where $[a,a^\dagger]=1$. We, therefore, arrive at 
\begin{align}\label{diffeq1}
 \left[ 2\alpha\hbar v_F ~ {a}^\dagger {a}-E^2 \right] \theta = 0~;~
 \left[ 2\alpha\hbar v_F ~ {a} {a}^\dagger-E^2 \right] \phi = 0,
\end{align}
from which we can immediately obtain the spinor part of the solution ${\tilde H}\psi=E\psi$ in terms of the eigenstates of the bosonic number operator $\hat{N}={a}^\dagger {a}$. The VP states for such a potential satisfy the energy quantization
\begin{align}\label{enquant1}
 E_N = {\rm sgn}(N)\sqrt{2\alpha\hbar v_F |N|}~;~N\in\mathbb{Z}.
\end{align}
We thus see that a linear potential, for massless Dirac particles, can lead to the genesis of discrete bound states at zero and non-zero energies. The zero-energy mode (for $N=0$), known as the Jackiw-Rebbi mode, is extensively studied, known to carry fractional quantum numbers \cite{goldstone1981fractional, lee2007edge}, and whose existence signals a topological phase transition for a sharp domain wall between two topologically distinct states. The finite-energy VP states corresponding to $N\neq 0$, on the other hand, are a salient feature of a Dirac oscillator \cite{moshinsky1989dirac, franco2013first, villalba1994exact} in a smoothly varying (here, linear) potential. We will later see the VP states foster nontrivial windings in their spin texture (unlike the Jackiw-Rebbi mode), which can be manipulated by applying an electric field.

The energy levels for the VP states in Eq.~\ref{enquant1} are not equidistant reminding us of the Landau quantization of Dirac fermions in graphene. The spatial profile of the VP states is obtained by solving
\begin{align}\label{diffeq3}
 \left[\hbar^2 v_F^2\partial_x^2 + (E^2 - \alpha^2x^2 + \alpha\hbar v_F) \right]\theta &= 0, \nonumber \\ 
 \left[\hbar^2 v_F^2\partial_x^2 + (E^2 - \alpha^2x^2 - \alpha\hbar v_F) \right]\phi &= 0.
\end{align}
The quantization in Eq.~\ref{enquant1} implies the normalized solution for the $N$-th bound state to be of the form (see Appendix~\ref{appA} for details)
\begin{align}\label{realspaceVP1}
 \theta_{|N|}(x) &= A_{|N|} e^{- x^2 /2\xi^2} H_{|N|}(x/\xi), \nonumber \\
 \phi_{|N|}(x) &= A_{|N|-1} e^{-x^2 /2\xi^2} H_{|N|-1}(x/\xi),
\end{align}
where $A_{|N|}^{-2}=2^{|N|}|N|!\pi^{1/2}\xi$ and $\xi=\sqrt{\hbar v_F/\alpha}$ ($H_{|N|}$ are the Hermite polynomials). The parameter $\xi$ is analogous to the magnetic length in the Landau level problem (identifying $\alpha/v_F$ with the magnetic field $eB$) over which the bound state wavefunctions localize along the edge. 

From the explicit forms of the wavefunctions (Eq.~\ref{realspaceVP1}), we can straightaway calculate observables like the probability and spin density associated with the VP spinor $|\psi_{|N|}(x)\rangle \equiv\begin{bmatrix} \theta_{|N|}(x) & \phi_{|N|}(x) \end{bmatrix}^T/\sqrt{2}$ keeping in mind that $\phi_{|N|}(x)$ does not contribute to the zero mode. This has a natural interpretation in terms of supersymmetry. The spectrum of $\theta$ and $\phi$ for finite energies are identical but the zero mode (for $\theta$) -- the former has one while the latter does not as is the case of the supersymmetric quantum mechanics with Witten index $\nu=1$. 

\begin{figure}
 \centering
  \includegraphics[width=\columnwidth]{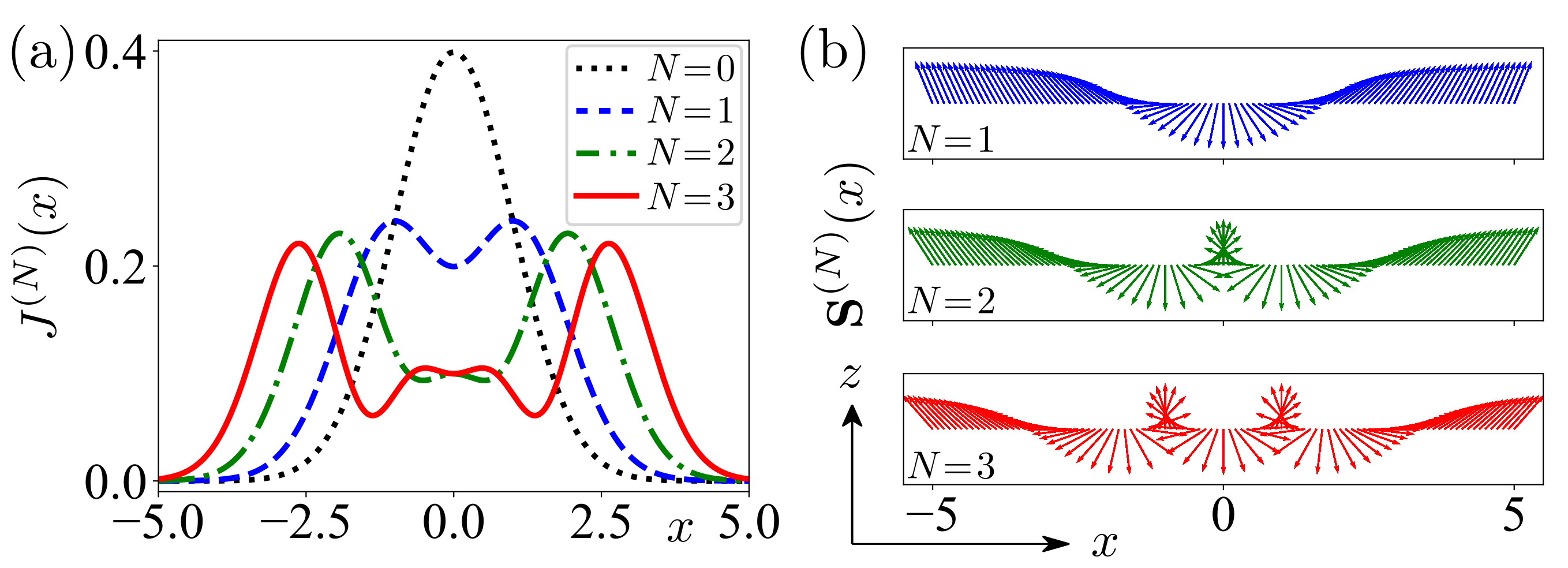}
  \caption{(a) The probability density $J^{(|N|)}(x)$ for $N=0$ to $3$ at $\alpha=0.5$. All states display a Gaussian decay away from $x=0$ where the magnetic field vanishes (the gap closing point). The VP states ($N\neq 0$) further show periodic modulations due to the Hermite polynomials in Eq.~\ref{realspaceVP1}. (b) Spatial profile of the unit spin vector ${\bf S}^{(|N|)}(x)$ for $N=1,2,3$ with its winding given by $N$. ${\bf S}^{(|N|)}(x)$ aligns along the $z$-axis in the asymptotic limit $x\rightarrow\pm\infty$.}
 \label{fig3}
\end{figure}

The probability density of a VP state with quantum number $\pm N$ is $J^{(|N|)}(x)=|\psi_{|N|}(x)|^2$ which we plot in Fig.~\ref{fig3} (a) at $\alpha=0.5$. For these wavefunctions involve a Gaussian component that decays over space (Eq.~\ref{realspaceVP1}), the corresponding probability density $J^{(|N|)}(x)$ dies off away from the gap closing point at $x=0$ with a characteristic length scale $\xi$. However, we uncover an interesting signature when we measure the spin density $S_i^{(|N|)}(x)=\langle \psi_{|N|}(x)|\sigma_i| \psi_{|N|}(x)\rangle$ where $\sigma_i$ are the Pauli matrices. For illustration, in Fig.~\ref{fig3} (b), we display the spatial profile of the normalized spin vector ${\bf S}^{(|N|)}(x)=(S_x^{(|N|)},0,S_z^{(|N|)})/\sqrt{[S_x^{(|N|)}]^2+[S_z^{(|N|)}]^2}$ for $N=1,2,3$ at $\alpha=0.5$ ($S_y^{(|N|)}=0$ as $\theta_{|N|}, \phi_{|N|}$ in Eq.~\ref{realspaceVP1} are real). Notably, Fig.~\ref{fig3} reveals a neat result: {\it The quantum number $N$ of the spinor $|\psi_{|N|}(x)\rangle$ measures its winding} along the extent of the helical edge over which the associated wavefunctions localize. The winding number for the $N$-th VP state is given by 
\begin{align}
    w^{(|N|)} = \frac{1}{2\pi} \int_{-\infty}^{+\infty} {\rm d}x~\partial_x{\rm arg}[S_z^{(|N|)}+iS_x^{(|N|)}].
\end{align}
As one asymptotically approaches $x\rightarrow\pm \infty$, the spin vectors associated with the VP spinors align or anti-align along the $z$-axis (depending on the sign of $\alpha$).
Note this is an invariant formulated in real space rather than the Fourier space since we do not have a translation-invariant HES [translation symmetry broken by $M(x)$] for which the momentum $\bf k$ can serve as a good quantum number and casting the spinor on to a Bloch sphere can reveal the associated spin vector. Such kind of topological spin texture in known in context of a quantum anomalous Hall insulator set on a cylindrical geometry \cite{wu2014topological}. 

The Jackiw-Rebbi mode does not wind since the corresponding spin vector ${\bf S}^{(0)}(x)$ only has a $z$-component. In distinction, localized periodic windings of the associated spin vector ${\bf S}^{(|N|)}(x)$ are observed for any VP state with index $N$ (with $N\neq 0$) as one traces it along the HES. Reminiscent of helicoids in the real space, this is a noteworthy signature of the VP states while the centers of the windings for the $N$-th VP state are given by the locations of the zeros of the Hermite polynomial $H_{|N|-1}(x/\xi)$ and coincide with the local minima of the associated probability density. This is remarkable  since by tuning the strength of the magnetic field, thereby changing $\xi$, one can alter the local spin polarization ${\bf S}^{(|N|)}(x)$ and vary the distance between the winding centers of ${\bf S}^{(|N|)}$ for $N\ge 2$. Mapping from the {\it extended} real line ${\mathbb R}:[-\infty,+\infty]$ to the spin space ${\cal S}_2$ (surface of a unit sphere), we observe the $N$-th VP state encircles an equatorial plane (in ${\cal S}_2$) $N$ times accumulating a spin-Berry phase of $N\pi$. 

Let us now investigate the emergence of the VP states in a smoothly varying but bounded potential that is more realistic to appear in an actual physical junction. This alters the spatial profile of the VP states and the associated energy quantization, and further and more important, limits the maximum number of such states within the gap due to the magnetic potential as will be demonstrated below.

\subsection{Bounded $M(x)$, $\mu(x)=0$}\label{sectwoC}

Here also we will assume no electric field is present by setting $\mu(x)=0$ and model the smooth profile of the bounded magnetic potential by a tan-hyperbolic function, namely, $M(x)=M_0\tanh(x/L)$, which is also referred to as the Rosen-Morse potential in the literature, where $M_0$ denotes the strength of the potential and the characteristic length $L$ the saturation length beyond which $M(x)$ attains the saturation values $\pm M_0$ on either side of $x=0$. Such type of magnetic potential is of relevance to modeling the physical interfaces in heterojunction-setups between topologically distinct phases and has been studied in graphene-based systems \cite{liu2015snake}.

In presence of such a bounded potential, the spatial profiles of the spinor components $\theta$ and $\phi$ are modified as follows. Substituting $M(x)=M_0\tanh(x/L)$ in Eq.~\ref{hamsimple6} and introducing the variable $z=\tanh(x/L)$, we obtain     
\begin{align}\label{hamVPtanh3}
    \left[ (1-z^2) \frac{d^2}{dz^2} -2z \frac{d}{d z} + l(l+1) - \frac{\lambda^2}{1-z^2} \right] \theta = 0, \nonumber \\
    \left[ (1-z^2) \frac{d^2}{dz^2} -2z \frac{d}{d z} + l(l-1) - \frac{\lambda^2}{1-z^2} \right] \phi = 0,
\end{align}
where $l=M_0L/\hbar v_F=L/\zeta$, $\zeta$ being a characteristic length scale inversely proportional to the saturation value of the magnetic potential $M_0$ and $\lambda=L\sqrt{M_0^2-E^2}/\hbar v_F$. For the bound states, in order to express $\theta, \phi$ in terms of polynomials, $l, \lambda$ must be integers with $0\le \lambda \le l$. These conditions together imply a finite number of VP states to exist within the gap $[-M_0, M_0]$ which is determined by the integer $l$ while their energy quantization is dictated by $\lambda$. Defining another integer-valued index $N\equiv l-\lambda$, the energy quantization for the VP states, which are the solution of Eq.~\ref{hamVPtanh3}, for the hyperbolic potential reads (see Appendix~\ref{appA} for details)
\begin{align}\label{tanhenergy}
    E_N= {\rm sgn}(N) M_0 \left[ \frac{2|N|}{l} - \frac{N^2}{l^2} \right]^{1/2},
\end{align}
while the spinor solution for the $N$-th bound state is given by the associated Legendre polynomials 
\begin{align}
    \theta \sim P_l^\lambda(z)~~;~~\phi \sim P_{l-1}^\lambda(z).
\end{align}
The zero mode corresponds to $N=0$ and is of the form $\theta\sim {\rm sech}^l(x/L)$ while the VP states at the edge of the spectrum, i.e., $N=l$ are non-normalizable and referred to as the {\it half-bound states} \cite{kennedy2002low}. 

The localization of the VP states is governed by the length scale $L$. As far as observables like the probability density and spin density are concerned, their behavior qualitatively remains the same as the linear case $M(x)\propto x$ but, to note, unlike the linear case, the winding number for the $N$-th VP state is close to but not exactly $N$. This is because of the potential being bounded, the spinor does not perfectly line up along the $z$-axis as $x\rightarrow\pm\infty$ (for the linear gradient this was not the case since the potential was unbounded and $S_x^{(N)}\rightarrow 0$ as $x\rightarrow\pm\infty$). For the hyperbolic gradient case, $S_x^{(N)}$ for the $N$-th VP state attains a non-zero value as $x\rightarrow\pm\infty$ multiplied by sgn($x$). This finite value decreases with $N$  and so does the deviation from $N$. We will comment on this issue in detail later at the end of this section.

\subsection{Unbounded $M(x)$, $\mu(x)\neq0$}\label{sectwoD}

We now embark on studying the response of the VP states to an external electric field. Let us first discuss the fate of the VP states that arise in a linear magnetic potential, additionally subject to a uniform electric field $\cal E$ that leads to chemical potential $\mu(x)=-{\cal E}x$. The spatial profile of the spinor components $\theta$ and $\phi$, in this case, are obtained from the decoupled differential equations  
\begin{align}\label{diffeqE2}
 \left[\partial_x^2 + \frac{E^2}{\hbar^2 v_F^2}+{\cal A}({\cal E}) + \frac{2E{\cal E}}{\hbar^2 v_F^2}x - {\cal A}^2({\cal E})x^2 \right] \theta &= 0, \nonumber \\
 \left[\partial_x^2 + \frac{E^2}{\hbar^2 v_F^2}-{\cal A}({\cal E}) + \frac{2E{\cal E}}{\hbar^2 v_F^2}x - {\cal A}^2({\cal E})x^2 \right] \phi &= 0,
\end{align}
where ${\cal A}({\cal E})=\sqrt{\alpha^2-{\cal E}^2}/\hbar v_F$. Note the condition for the chemical potential to remain within the magnetic gap implies ${\cal E}\le \alpha$.

For bound states, we seek the solutions to be expressed in terms of polynomials which enforces a quantization of their energy (see Appendix~\ref{appA} for details)
\begin{align}\label{energyE1}
 E_N({\cal E}) = {\rm sgn}(N)\sqrt{2\alpha\hbar v_F |N|}\left(1-\frac{{\cal E}^2}{\alpha^2} \right)^{3/4},
\end{align} 
depending on the electric field strength ${\cal E}$. The spatial forms of $\theta$ and $\phi$ of the $N$-th bound state are given by 
\begin{align}\label{realspaceVPE1}
 \theta_{|N|}(x) &= B_{|N|} e^{- (x-x_0)^2 /2\xi_0^2} H_{|N|}[(x-x_0)/\xi_0], \nonumber \\
 \phi_{|N|}(x) &= B_{|N|-1} e^{- (x-x_0)^2 /2\xi_0^2} H_{|N|-1}[(x-x_0)/\xi_0],
\end{align}
where $B_{|N|}^{-2}=2^{|N|}|N|!\pi^{1/2}\xi_0$, $\xi_0=\sqrt{1/{\cal A}({\cal E})}$, and $x_0=E_{N}{\cal E}/(\alpha^2-{\cal E}^2)$.

\begin{figure}
 \centering
  \includegraphics[width=\columnwidth]{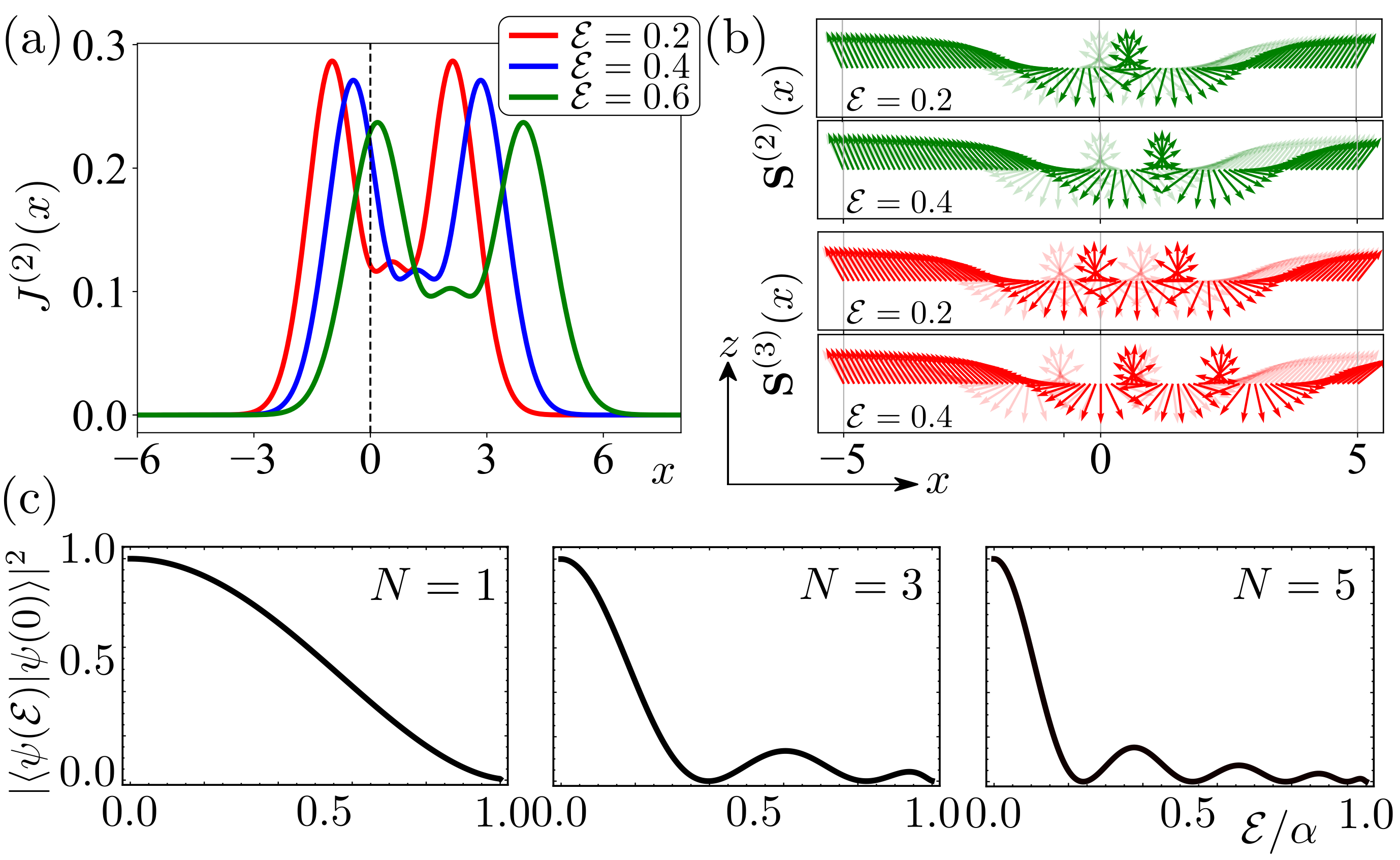}
  \caption{(a) The probability density $J^{(N)}(x)$ for $N=2$ at various values of the electric field strength ${\cal E}$ and $\alpha=0.8$. (b) Spatial profile of the unit spin vector ${\bf S}^{(N)}(x)$ for $N=2,3$ at different values of ${\cal E}$. A shift due to a finite ${\cal E}$ is noticeable compared to the ${\cal E}=0$ case (faint arrows) for which the plot of $J^{(N)}(x)$ and the profile of ${\bf S}^{(N)}(x)$ will be symmetric about $x=0$ (see Fig.~\ref{fig3}). (c) A plot of the spinor overlap function $|\langle\psi({\cal E})|\psi(0)\rangle|^2$ as a function of ${\cal E}/\alpha$ for the VP states with $N=1,3,5$ showing the evolution of the VP states starting from ${\cal E}=0$. Complete orthogonality for each $N$ is realized at the critical field ${\cal E}=\alpha$.}
 \label{fig5}
\end{figure}

The application of an electric field unfolds an intriguing interplay signaling a transition at a critical field ${\cal E}_0=\alpha$ that has striking signatures in the spectrum as well as the spatial profiles of the probability density $J(x)$ and the (normalized) spin density ${\bf S}(x)$.

Firstly, the energy spacing between the VP states decreases with the electric field ${\cal E}$ (as per the scaling noted in Eq.~\ref{energyE1}) as ${\cal E}\rightarrow \alpha$ and the entire tower collapses at the transition point ${\cal E}=\alpha$. This is a remarkable phenomenon from the viewpoint of supersymmetric quantum mechanics -- the associated Witten index ceases to qualify for a {\it topological invariant} of the full supersymmetric theory as the number of zero modes alters drastically at this critical point.

Secondly, when the chemical potential due to the applied electric field is within the gap $[-M_0, M_0]$ set by the magnetic potential $M(x)$, {\it i.e.}, ${\cal E}<\alpha$, the effect of the electric field manifests as a shift of the spatial profile of the observables such as the probability density and the spin texture of the VP states for $N\neq 0$ in  as we see in the plot of $J(x)$ in Fig.~\ref{fig5} (a). Here we have plotted $J(x)$ for the $N=2$ VP state at different values of ${\cal E}$. Evidently, the shift $x_0$ increases with ${\cal E}$ but so does the localization length $\xi_0$; and as a result, the magnitude of $J(x)$ is diminished. The Jackiw-Rebbi mode remains unaffected by the electric field since the shift parameter $x_0$ vanishes for the zero mode.  

The electric field also influences the winding pattern of the spin vector associated with each VP state. The locations of the winding centers are shifted along the direction of the electric field while the separation between two successive winding centers decreases with the electric field. This is shown in Fig.~\ref{fig5} (b) for the VP states with $N=2,3$ at different values of ${\cal E}$. Because of such a shift of the winding structure associated with the eigenstates $\psi$, when we compute the overlap of a given eigenstate, indexed by $N$, at a finite value of $\cal E$ with that for ${\cal E}=0$, we observe a periodic modulation in the profile of ${\cal O}({\cal E})\equiv |\langle\psi({\cal E})|\psi(0)\rangle|^2$ as a function of ${\cal E}/\alpha$. Notably, each VP state at the critical value of the electric field ${\cal E}=\alpha$, when they are the part of a degenerate manifold of zero modes, becomes orthogonal to itself at ${\cal E}=0$. This is visible in Fig.~\ref{fig5} (c). The zeros of the overlap function ${\cal O}({\cal E})$ are given by the values of $\cal E$ at which the shifted spin texture is orthogonal to that at ${\cal E}=0$. The number of such zeros for the $N$-th VP state is, therefore, $N$ for it has $N$ winding centers.

We will now investigate the effect of a spatially textured electric field ${\cal E}(x)$ such that the resultant chemical potential $\mu(x)\propto M(x)$ with the condition $|\mu(x)| \le \mu_0$, $\pm\mu_0$ being the asymptotic values of $\mu(x)$ at $x\rightarrow \pm \infty$, $|M(x)| \le M_0$, $\pm M_0$ being the asymptotic values of $M(x)$ at $x\rightarrow \pm \infty$, and $\mu_0\le M_0$.

\subsection{Bounded $M(x)$, $\mu(x)\neq 0$}\label{sectwoE}

\begin{figure}
 \centering
  \includegraphics[width=\columnwidth]{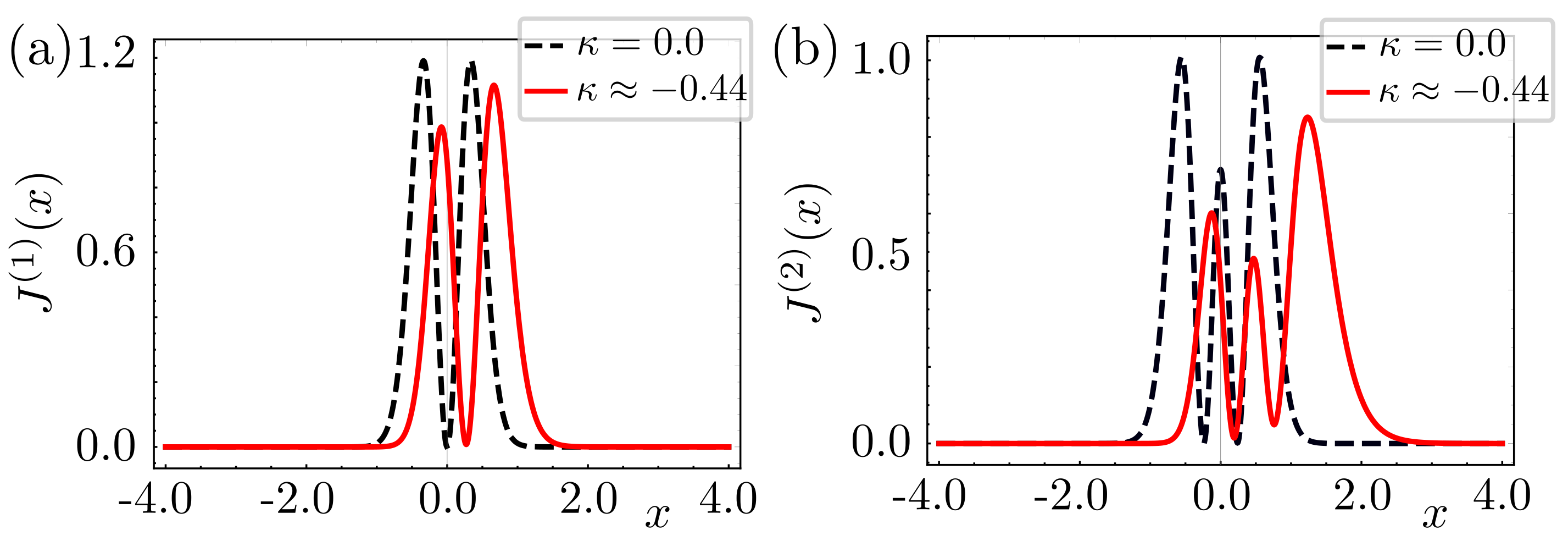}
  \caption{(a) The probability density for the first VP state $J^{(1)}(x)$ for a hyperbolic profile of the magnetic potential $M(x)=10\tanh(x)$ and in the presence of an electric-field induced chemical potential $\mu(x)=10\kappa\tanh(x)$ at various values of the electric field strength $\kappa$. (b) The same for the second VP state $J^{(2)}(x)$. The black dashed line is the case for no electric field $\kappa=0$ presented as a reference with respect to which a shift in $J^{(N)}(x)$ is clearly visible (while the maxima are diminished).}
 \label{fig6}
\end{figure}

\begin{figure}
 \centering
  \includegraphics[width=\columnwidth]{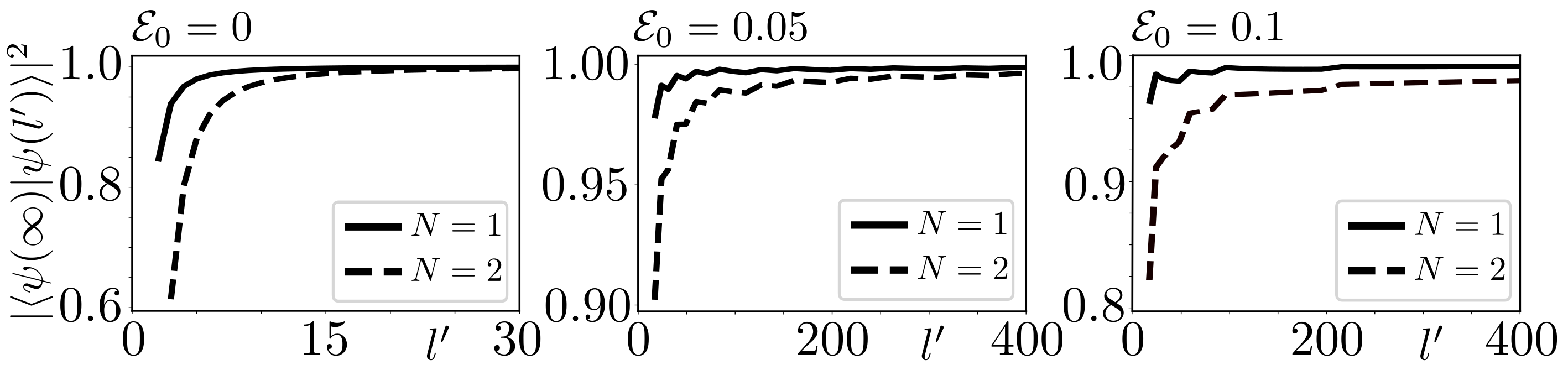}
  \caption{Shown is the behavior of the overlap function $ |\langle\psi(\infty)|\psi(l')\rangle|^2$ for the VP states with $N=1,2$ against the number $l'$ (where $2l'+1$ is the number of bound states present in the system) at various values of the electric field strength ${\cal E}_0=-\kappa M_0/L$. The overlap approaches 1 asymptotically as $l'\rightarrow\infty$, although the growth gets slower with ${\cal E}_0$. Here $l'$ is an integer increased in steps of 1  keeping $M_0/L$ fixed while changing $L$.}
 \label{fig8}
\end{figure}

A simple setting where the above conditions for the chemical potential $\mu(x)$ are satisfied is to consider $\mu(x)=\kappa M(x)=\kappa M_0\tanh(x/L)$ with $|\kappa|\le 1$. The decoupled equations for the spinor components $\theta$ and $\phi$ then read
\begin{align}\label{electrictanh1}
 \left[-\partial_x^2 + W^2(x) - \partial_x W(x) \right] \theta &= \varepsilon \theta,  \nonumber \\
 \left[-\partial_x^2 + W^2(x) + \partial_x W(x) \right] \phi &= \varepsilon \phi,
\end{align}
where 
\begin{align}\label{electrictanh2}
 W(x)= \frac{\sqrt{1-\kappa^2}}{\hbar v_F} \left[M(x)+\frac{\kappa}{{1-\kappa^2}} E\right],
\end{align}
and 
\begin{align}\label{supereigenvalue}
 \varepsilon = E^2/\left[\hbar^2v_F^2(1-\kappa^2)\right].
\end{align}
Eq.~\ref{electrictanh1} is reminiscent of the supersymmetry problem posed in the beginning in Eq.~\ref{hamsimple6}. Here we will elicit the concept of shape invariance \cite{gangopadhyaya2017supersymmetric} of the supersymmetric partner potentials $W^2 \pm\partial_x W$ to analytically compute the bound state spectrum since the superpotential $W(x)$ can be cast to the Rosen-Morse form $W(x)=A\tanh(x/L)+B/A$ \cite{gangopadhyaya2017supersymmetric, dabrowska1988explicit}. The spectrum of the VP states, in this case, turns out to be (see Appendix~\ref{appB} for details)
\begin{align}\label{energytanhelectric}
 E_N = {\rm sgn}(N) \sqrt{\frac{M_0^2(1-\kappa^2) - (M_0 \sqrt{1-\kappa^2} - \hbar v_F |N|/L)^2}{1+ M_0^2 \kappa^2/(M_0 \sqrt{1-\kappa^2} - \hbar v_F |N|/L)^4 }}. 
\end{align}
We can immediately retrieve the spectrum for the zero electric field case (Eq.~\ref{tanhenergy}) upon setting $\kappa=0$ in the above equation. 

To obtain the spatial profile of the VP states, we cast Eq.~\ref{electrictanh1} as
\begin{align}\label{electrictanh2}
(1-z^2) \frac{d^2 \theta}{dz^2} -2z \frac{d \theta}{dz} + \left[l'(l'+1) - \frac{\lambda'^2}{1-z^2} \right] \theta &= 0, \nonumber \\
(1-z^2) \frac{d^2 \phi}{dz^2} -2z \frac{d \phi}{dz} + \left[l'(l'-1) - \frac{\lambda'^2}{1-z^2} \right] \phi &= 0,
\end{align}
where the variable $z=\tanh(x/L)$, $l'=M_0L\sqrt{1-\kappa^2}/\hbar v_F$, and $\lambda'=L\sqrt{(1-\kappa^2)M_0^2-E^2+2E\kappa M_0 z}/\hbar v_F$. We identify the parameters $l', \lambda'$ with $l, \lambda$ for the zero electric field case when $\kappa=0$. The solutions of Eq.~\ref{electrictanh2} are given in terms of Jacobi polynomials when $l'$ is an integer that physically represents the number of VP states present in the system. Thus we obtain the spatial profile of the spinor components $\theta$ and $\phi$ of the $N$-th bound state as
\begin{align}
 \theta_{|N|}(z) &\sim \left(1-z\right)^{a_{|N|}/2}\left(1+z\right)^{b_{|N|}/2}{\cal P}^{a_{|N|},b_{|N|}}_{|N|}(z), \nonumber \\
 \phi_{|N|}(z) &\sim \left(1-z\right)^{a'_{|N|}/2}\left(1+z\right)^{b'_{|N|}/2}{\cal P}^{a'_{|N|},b'_{|N|}}_{|N|}(z),
\end{align}
where ${\cal P}_N^{(a,b)}$ are the Jacobi polynomials, $a_N=l'-N+l'LE\kappa/\left[\hbar v_F(l'-N)\sqrt{1-\kappa^2}\right]$, $b_N=2(l'-N)-a_N$, $a'_N=a_N(l'\rightarrow l'-1)$, $b'_N=2(l'-N-1)-a'_N$. Here the index $N$ is bounded by $|N|\le l'$ and the zero mode ($N=0$) exists as a solution only for $\theta$. Therefore, the total number of non-zero energy bound states in the system amounts to $2l'$.

With this, we can compute the probability density $J^{(N)}(x)$ which we plot in Fig.~\ref{fig6}. For the plot, we consider $M_0/\hbar v_F=10$ and $L=1$ setting the maximum number of VP states allowed in the system to $2l'=20$ when no electric field is applied {\it i.e.}, $\kappa=0$. We then apply an electric field specified with the dimensionless parameter $\kappa\approx 0.44$ that reduces $l'$ to $9$, or equivalently, $2l'=18$. We plot $J^{(N)}(x)$ for $N=1$ [Fig.~\ref{fig6} (a)] and $N=2$ [Fig.~\ref{fig6} (b)] both evincing the spatial shift due to a finite electric field and bounded chemical potential.   

Finally, we would like to highlight how the VP wavefunctions for the bounded case {\it i.e.}, with a finite number of bound states ($l'$) differ from the unbounded ones by computing their overlap for a given index $N$, both in presence and absence of the electric field. This is shown in Fig.~\ref{fig8} for $N=1, 2$. That the wavefunctions for the bounded case lack an integer-valued winding number to characterize their spin texture is evident from their overlap with the unbounded case deviating from 1. The overlap function ${\cal O}(l')\equiv |\langle\psi(\infty)|\psi(l')\rangle|^2$ asymptotically approaches 1 as $l'\rightarrow \infty$, however, the growth of ${\cal O}(l')$ towards 1 gets progressively slower with the electric field  as noticeable in Fig.~\ref{fig8}. Since $l'$ is directly proportional to the mass gap $M_0$, this supports our claim that when the mass gap becomes quite large, the spin texture of the VP states for the bounded case almost attains complete windings determined by their quantization index $N$.

\section{Detecting the VP states in a lattice model simulation}\label{secthree}

\begin{figure}
 \centering
  \includegraphics[width=\columnwidth]{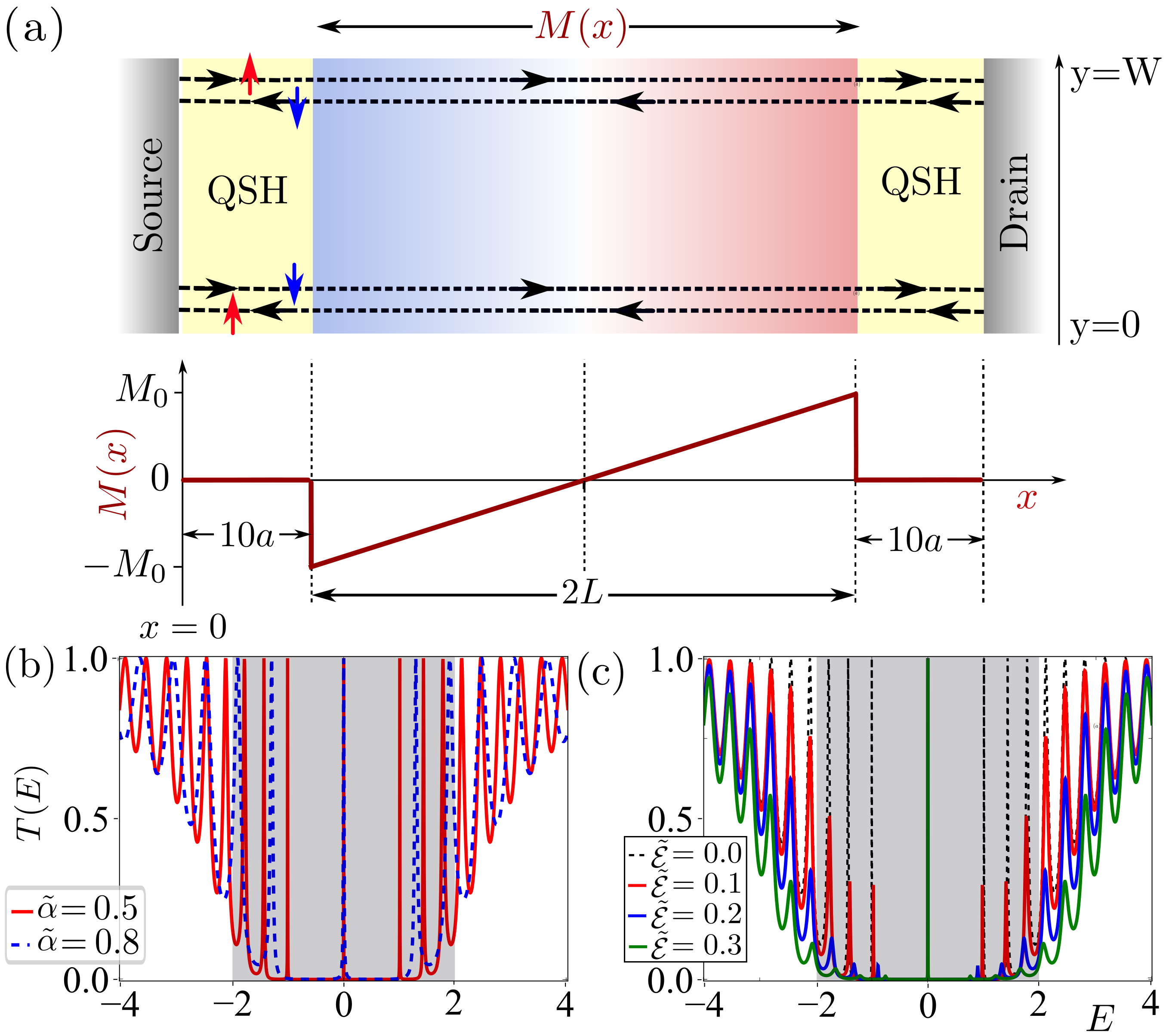}
  \caption{(a) The schematic of the lattice model used for KWANT simulation. From source to drain connects the QSH region subjected to a magnetic potential $M(x)$ that varies linearly from $-M_0$ to $M_0$ over an extent of $2L$ (the color-graded region in the middle). The transmission probability $T(E)$ is measured as a function of the incident energy of the electrons $E$ and in the units of the conductance quantum $G_0=2e^2/h$. Resonances are observed when $E$ coincides with the bound state energies. Shown in (b) are the resonances for various values of the ramp parameter $\tilde{\alpha}\equiv\alpha\hbar v_F$ when no electric field is present. The energy spacing of the neighboring VP states  $\Delta E$ increases with $\tilde{\alpha}$ following $\Delta E\sim \sqrt{\tilde{\alpha}}$. (c) The same in the presence of a uniform electric field of strength ${\cal E} < \alpha$. (the effective electric field is denoted as $\tilde{\cal E}={\cal E}\hbar v_F$). Here $\Delta E$ decreases with ${\cal E}$ with the peaks of the resonances gradually diminishing.}
 \label{fig7}
\end{figure}

To demonstrate the emergence of the VP states in a more realistic setup, we consider transport simulations performed on the lattice model of a two-dimensional topological insulator hosting one-dimensional helical edge states at its boundary, using the package KWANT \cite{groth2014kwant}. The energy-resolved transmission probability renders evidence for the existence of the bound states (the VP states) and demonstrates the parametric dependence of their spectrum on the magnetic field and the electric field, which is in tune with the results obtained from the analytical results of one-dimensional helical edge model discussed in Section \ref{sectwo}. Specifically, here we will be considering the case of a linearly varying magnetic field and a uniform electric field.   

We consider a two-dimensional square lattice of rectangular geometry with a length of $2(L+10a)$ and width $W=200a$ [Fig.~\ref{fig7} (a)], where $a$ is the lattice spacing (explicit values are mentioned later). A Zeeman field is applied along the length of the rectangular sample which is along the $x$-axis,  such that the strength of the field is given by $M(x,y)=\alpha (x-x_c)$, $x_c$ being the center of the rectangle situated at $x_c=L+10a$. The parameter $\alpha$ is fixed by a cutoff $M_0$, which lies well within the bulk gap, such that $M_0=\alpha L$.

Let us now specify the model which is simulated using the KWANT package. The bulk, in the pristine form, is described by the BHZ Hamiltonian for the HgTe/CdTe quantum well \cite{bernevig2006quantum}
\begin{align}
H_{\rm BHZ} = -D k^2 + A k_x \sigma_z \tilde{\sigma}_x - A k_y \tilde{\sigma}_y + ({\cal M}-B k^2) \tilde{\sigma}_z,
 \label{BHZham}
\end{align}
where $\sigma$ and $\tilde{\sigma}$ denote the Pauli matrices to describe the spins (up or down along $S_z$) and the orbitals (electron type or hole type) respectively and $A$, $B$, $D$ and ${\cal M}$ are material dependent parameters. We discretize the Hamiltonian in Eq.~\ref{BHZham} to obtain a tight-binding version on our square lattice but with a basis of two sites (to represent the two-level system of orbitals). Using $k^2 = 2 a^{-2}[2-\cos(k_x a)-\cos(k_y a)]$, $k_x = a^{-1} \sin(k_x a)$, and $k_y = a^{-1} \sin(k_y a)$ the tight-binding Hamiltonian reads
\begin{equation}
H_{\rm tb} = \sum_i (c_i^{\dagger} H_{i,i+a_x} c_{i+a_x} + c_i^{\dagger} H_{i,i+a_y} c_{i+a_y} + {\rm h.c.}) + c_i^{\dagger} H_{ii} c_i,
\end{equation}
where $c_i^{\dagger} \equiv (c^{\dagger}_{i,s,\uparrow},c^{\dagger}_{i,p,\uparrow},c^{\dagger}_{i,s,\downarrow},c^{\dagger}_{i,p,\downarrow})$ denotes the set of creation operators for the electrons in $s$ and $p$ orbital with $\uparrow$ and $\downarrow$ spins at site $i$ with coordinates $i=(i_x,i_y)$; $a_x = a(1,0)$ and $a_y = a(0,1)$ are the lattice vectors with $a$ being the lattice constant. Each of the terms, $H_{ii}$ and $H_{i,i+a_x(a_y)}$, is a $4 \times 4$ block matrices defined by
\begin{align}
H_{ii} &= -\frac{4 D}{a^2} - \frac{4 B}{a^2} \tilde{\sigma}_z + {\cal M} \tilde{\sigma}_z, \nonumber \\
H_{i,i+a_x} &= \frac{D + B \tilde{\sigma}_z}{a^2} + \frac{A \sigma_z \tilde{\sigma}_x}{2 i a}, \nonumber \\
H_{i,i+a_y} &= \frac{D + B \tilde{\sigma}_z}{a^2} + \frac{i A \tilde{\sigma}_y}{2 a}.
\label{bhzlattice}
\end{align}
In the region where the spatially varying in-plane magnetic field and a uniform electric field are applied along the $x$-direction, the diagonal term in Eq.~\ref{bhzlattice} is modified to 
\begin{align}
 H_{ii} = -\frac{4 D}{a^2} - \frac{4 B}{a^2} \tilde{\sigma}_z + {\cal M} \tilde{\sigma}_z + M_i \sigma_x + {\cal E}(x_i - x_c),
\label{BHZlatgrad}
\end{align}
where $M_i=\alpha (x_i - x_c)$, $x_i$ being the $x$-coordinate of the $i$-th lattice site and ${\cal E}$ the strength of the electric field. The last term in Eq.~\ref{BHZlatgrad} is the discretized form of the chemical potential $\mu(x,y)=-{\cal E}(x-x_c)$ which is present all over the lattice owing to the electric field.

The standard parameters for the HgTe/CdTe quantum wells that are used in Eq.~\ref{bhzlattice} are~\cite{konig2007quantum} $A = \hbar v_F = 364.5$ nm meV, $B = -686$ ${\rm nm}^2$ meV, and ${\cal M} = -15$ meV while $D$ is set to zero to place the Dirac cone at zero energy, and the lattice constant $a=3$ nm. The length $L$ over which the magnetic and the electric field are operating is decided by setting the parameter $\alpha=M_0/L$ to different values (here we consider $\alpha\hbar v_F=0.5$ and $\alpha \hbar v_F=0.8$ which yield $L=486a$ and $L=303a$ respectively) with $M_0=2$ meV, much less than the bulk gap determined by ${\cal M}$. The magnetic field is then ramped from $-M_0$ to $M_0$ over a length of $2L$ [Fig.~\ref{fig7} (a)].  When present, the values of the electric field are chosen such that ${\cal E}<\alpha$. 

With the essentials of our setup provided, we turn to measure observables such as the transmission probability $T$ as a function of the incidence energy $E$. Such a plot will feature resonances from which the existence of the VP states can be readily verified. We do so at distinct values of the magnetic ramp parameter $\alpha$ and the electric field ${\cal E}$ to elucidate the parametric dependence of the energy spectrum of these bound states. 

The plot in Fig.~\ref{fig7} (b) displays the behavior of $T(E)$ at different values of $\alpha$ (with ${\cal E}=0$) from which we observe the VP states appearing at quantized values of $E$ with sharp resonance peaks when a linearly varying magnetic field is forming a smooth interface between two topologically distinct regions identified with sgn$[M(x)]$. The Jackiw-Rebbi zero mode is visible at $E=0$. The energy spacing between neighboring bound states increases with $\alpha$ following Eq.~\ref{enquant1}. With $\alpha$ taking higher values, the interface becomes sharper, and as a result, the VP states gradually disappear in the bulk with the Jackiw-Rebbi zero mode pinned to $E=0$. 

When a uniform electric field of strength ${\cal E}$ is applied on top of it, the expression for the bound state energy spectrum is modified according to Eq.~\ref{energyE1}. A reverse phenomenon is now observed. The energy spacing between the neighboring bound states gets shrunk with ${\cal E}$ and the height of the resonance peaks diminished as visible in Fig.~\ref{fig7} (c). Here we have set $\alpha \hbar v_F =0.5$ and considered various values of $\cal E$ such that ${\cal E}<\alpha$ or equivalently, ${\cal E}\hbar v_F < 0.5$. The bound states start collapsing onto the Jackiw-Rebbi mode as ${\cal E}\rightarrow \alpha$ and exactly at ${\cal E}=\alpha$ they form a highly degenerate manifold of zero modes. Together, these results illuminate how one can engineer such finite-energy sub-gap states at the interface of distinct topological states via the elegant interplay of an in-plane electric field and the spatially-textured magnetic field generating the interface.

\section{Conclusion}\label{secfour}  

We first consider the problem of a linearly varying mass gap for the helical edge states of a quantum spin Hall insulator. This reduces to the problem of a Dirac oscillator in 1D and leads to the appearance of bound states at zero as well as finite energies, the latter identified with the Volkov-Pankratov states. The eponymous bound states, indexed with an integer $N$, have a remarkable characteristic feature -- the real-space spin texture of the associated eigenfunctions has topological windings given by $N$. We further show that the application of an electric field along the edge enables control of the position of the winding centers as well as it results in a nontrivial scaling of the energy spacing between successive bound states. As the linear potential is unbounded, it is unrealistic as far as an experimental realization is concerned. We therefore also consider the problem of a hyperbolic potential that saturates to finite values at large distances. We show for such potential, a finite number of sub-gap states appear whose count is determined by the parameters of the underlying potential. The nontrivial spin textures continue to exist for this case, however, a bounded potential is found to hinder associating an integer-valued number to the full winding pattern. Finally, we show that the application of an electric field leads to a total collapse of the bound state spectrum onto the zero mode when tuned to a critical value.

\section{Acknowledgments}
We thank Dibyakanti Mukherjee for discussions during the initial stages of the work. VA acknowledges support from IISER Kolkata in the form of a subsistence grant. SC acknowledges the Council of Scientific and Industrial Research (CSIR), Govt. of India for financial
support in the form of a fellowship. KR thanks the sponsorship, in part, by the Swedish Research Council. SD would like to acknowledge the MATRICS grant (MTR/ 2019/001 043) from the Science and Engineering Research Board (SERB) for funding. We acknowledge the central computing facility (DIRAC supercomputer) and the computational facility at the Department of Physics (KEPLER) at IISER Kolkata. \\

{\it Author contribution}: The first two authors, VA and
SC have contributed equally to this work.

\bibliographystyle{apsrev}
\bibliography{references}


\appendix

\section{Bound state spectrum of the VP states}\label{appA}

In this appendix, we will show how to derive the bound state spectrum of the VP states from the associated wavefunctions. The four distinct cases are as follows. 

\subsection{Unbounded $M(x)$, $\mu(x) = 0$ }

The decoupled equations for $\theta$ and $\phi$, as given in Eq.~\ref{diffeq3}, read
\begin{equation}
    \frac{d^2 y}{d x^2} - k_1x^2y + k_2y =0,
\end{equation}
where, $k_1 = \alpha^2/\left(\hbar^2 v_F^2\right)$ and $k_2 = \left(E^2 + \sigma \alpha \hbar v_F\right)/\left(\hbar^2 v_F^2\right)$ with $\sigma = +1, -1$ for $y=\theta$ and $y=\phi$ respectively. A general form of the solution to the above equation is
\begin{equation}
y(x) \sim e^{-\left(\sqrt{k_1}/2\right) x^2} H_{\frac{1}{2} \left(k_2/\sqrt{k_1} -1\right)} \left( k_1^{1/4} x\right).
\end{equation}
A polynomial form of the solution (representing a bound state) then demands the index of the Hermite polynomial is an integer implying
\begin{equation}
    \frac{1}{2}\left( \frac{k_2}{\sqrt{k_1}} -1\right) = |N| \implies E = {\rm sgn}(N)\sqrt{2 \alpha \hbar v_F |N|},
\end{equation}
yielding the spectrum of the bound states as noted in Eq.~\ref{enquant1}.

\subsection{Bounded $M(x)$, $\mu(x) = 0$}

The solutions of the differential equations in Eq.~\ref{hamVPtanh3} are given by the associated Legendre polynomials: $\theta \sim P_l^{\lambda}(z)$ and $\phi \sim P_{l-1}^{\lambda}(z)$, where $z=\tanh\left(x/L\right)$, $l = M_0 L/\hbar v_F$, and $\lambda=L\sqrt{M_0^2-E^2}/\hbar v_F$. In order to have bound state solutions, the condition to be satisfied is that $l$ and $\lambda$ take integer values while $\lambda \leq l$ for $\theta$ and $\lambda \leq l-1$ for $\phi$. Therefore, we can write
\begin{equation}
\begin{split}
    & \lambda = l - |N|\\
    \implies & E = {\rm sgn}(N) M_0 \left[ \frac{2|N|}{l} - \frac{N^2}{l^2} \right]^{1/2}.
\end{split}
\end{equation}
For $N = 0$, we obtain $\lambda = l$ and $E = 0$ which corresponds to the zero mode whose spatial profile is dictated by $P_l^l(z) \sim (1-z^2)^{l/2}  = \sech^l(x/L)$. On the other hand, for $\lambda = 0$, we obtain the half-bound state solutions at the edge of the spectrum $E=\pm M_0$ \cite{kennedy2002low}. 

\subsection{Unbounded $M(x)$, $\mu(x)$ $\neq 0$}

Eq~\ref{diffeqE2} can be expressed as
\begin{equation}
    \frac{d^2 y}{d x^2} + (-ax^2+bx+c)y=0,
\end{equation}
where
\begin{equation}
	\begin{split}
		a &= \left( \frac{\alpha^2}{\hbar^2 v_F^2} - \frac{{\cal E}^2}{\hbar^2 v_F^2} \right),\\
		b &= \frac{2E {\cal E}}{\hbar^2 v_F^2},\\
		c &= \left(\frac{E^2}{\hbar^2 v_F^2} + \sigma \sqrt{\frac{\alpha^2}{\hbar^2 v_F^2}-\frac{{\cal E}^2}{\hbar^2 v_F^2}}\right),
	\end{split}
\end{equation}
and $ \sigma = +1, -1$ for $y=\theta$ and $y=\phi$ respectively. The solution of the above equation is given by
\begin{equation}
    y(x) \sim e^{-\left[\frac{(2ax-b)}{\sqrt{2} a^{3/4}}\right]^2/4 } H_{\frac{b^2-4a^{3/2} + 4ac}{8a^{3/2}}} \left(\frac{2ax-b}{2 a^{3/4}} \right).
\end{equation}
Therefore, applying the condition that the index of the Hermite polynomial will be a non-negative integer for a bound state solution, we find
\begin{align}
    &~~~~~~~~~~~\frac{b^2-4a^{3/2} + 4ac}{8a^{3/2}} = |N| \nonumber \\
    &\implies E = \sgn(N) \sqrt{2 \alpha \hbar v_F |N|}\left(1-\frac{{\cal E}^2}{\alpha^2} \right)^{3/4},
\end{align}
as noted in Eq.~\ref{energyE1}.

\section{Bound state spectrum from supersymmetric shape invariance}\label{appB}

For details on the shape invariance of solvable supersymmetric potentials, the reader is referred to Ref.~ \onlinecite{gangopadhyaya2017supersymmetric}. Here we provide a sketch of the underlying concept and its illustration with examples relevant to the QSHS problem at hand.

In quantum mechanics, a pair of systems described by the Hamiltonians
\begin{align}
 {\cal H}_{\pm} = -\frac{d^2}{dx^2} + V_\pm(x,a),
\end{align}
with the parameter $a$ specifying the potentials, are called supersymmetric partners of each other with an identical spectrum (up to zero modes). The two potentials, called the partner potentials, $V_\pm(x,a)$ derive from a superpotential $W(x,a)$ as
\begin{align}
  V_\pm(x,a) =  W^2(x,a) \pm \frac{dW}{dx}.
\end{align}
The supersymmetric shape invariance condition applies to a specific class of superpotentials $W(x,a)$ such that the two (shape invariant) partner potentials are related by
\begin{align}\label{susyqmpot1}
 V_+(x,a_0) - V_-(x,a_1) =  g(a_1) - g(a_0),
\end{align}
where $g(a_1) - g(a_0) $ is an additive constant and $a_1=f(a_0)$, some function of $a_0$, generalizing to $a_{i+1}=f(a_{i})$. Note the coordinate dependence in $V_+(x,a_0)$ and $V_-(x,a_1)$ is the same but the parameter values are different.

At the level of the spectra for the two systems, characterized by the energy eigenvalues $E^+_n$ and $E^-_n$ for ${\cal H}_+$ and ${\cal H}_-$ respectively, Eq.~\ref{susyqmpot1} implies
\begin{align}\label{shapeinveval1}
 E^+_n(a_0) - E^-_n(a_1) =  g(a_1) - g(a_0),~~ \forall n
\end{align}
with $E_0^-(a_i)=0,~i=0,1,\dots,n$ for an unbroken supersymmetry. With the method of induction, it is straightforward to show that 
\begin{align}\label{shapeinveval2}
 E_n^-(a_0) = g(a_n) - g(a_0),
\end{align}
and the supersymmetry demands $E_n^+(a_i)=E_{n+1}^-(a_i)$. Therefore, knowing $W(x,a)$ and identifying $a_0$, $a_1$, and $g(a_0)$ suffice to compute the entire spectrum of ${\cal H}_\pm$.

For example, if we try to compute the spectrum of ${\cal H}_-(x,a_0)$, the starting point is to note $E^-_0(a_0)=0$. Then the next eigenvalue will be given by
\begin{align}
 E^-_1(a_0) &= E^+_0(a_0) \nonumber \\
 &= E^-_0(a_1) + g(a_1) - g(a_0) \nonumber \\
 &= g(a_1) - g(a_0),
\end{align}
(using Eq.~\ref{shapeinveval1}). Similarly,
\begin{align}
 E^-_2(a_0) &= E^+_1(a_0) \nonumber \\
 &= E^-_1(a_1) + g(a_1) - g(a_0) \nonumber \\
 &= E^+_0(a_1) + g(a_1) - g(a_0) \nonumber \\
 &= E^-_0(a_2) + g(a_2) - g(a_1) + g(a_1) - g(a_0) \nonumber \\
 &= g(a_2) - g(a_0).
\end{align}
Continuing this way, we end up at Eq.~\ref{shapeinveval2}.

Let us now illustrate how we can use the concept of shape invariance to compute the spectrum of the VP states in presence of an electric field. The following two cases are of concern.

\subsection{Unbounded $M(x)$, $\mu(x)$ $\neq 0$}

Here we consider $M(x)=\alpha x$ and $\mu(x)=-{\cal E}x$. 
Eq.~\ref{diffeqE2} can be expressed as 
\begin{align}
 \left[-\partial_x^2 + W^2(x) - \partial_x W(x) \right] \theta &= \varepsilon \theta,  \nonumber \\
 \left[-\partial_x^2 + W^2(x) + \partial_x W(x) \right] \phi &= \varepsilon \phi,
\end{align}
where $W(x)=Ax+B$ with
\begin{align}
 A=\sqrt{\alpha^2-{\cal E}^2}/\hbar v_F\,,~ B=E{\cal E}/\left(\hbar^2v_F^2 A \right),
\end{align}
and 
\begin{align}\label{superpotentialen1}
 \varepsilon = E^2\alpha^2/\left[\hbar^2v_F^2(\alpha^2-{\cal E}^2)\right].
\end{align}
It is straightforward to identify $g(a_i)=2Aa_i$, $a_i=a_0+i$, $i=0,1,\dots,|N|$, and so,
\begin{align}
 \varepsilon = 2A|N|,
\end{align}
from which Eq.~\ref{energyE1} follows. The case for $\mu(x)=0$ is obtained simply by setting ${\cal E}=0$ is Eq.~\ref{superpotentialen1}.

\subsection{Bounded $M(x)$, $\mu(x)$ $\neq 0$}

In this case, the superpotential is of the form
\begin{align}
    W(x) = A\tanh{(x/L)} + B
\end{align}
with
\begin{align}
    A = \frac{\sqrt{1-\kappa^2} M_0}{\hbar v_F}~;~
    B = \frac{\kappa E}{\hbar v_F \sqrt{1-\kappa^2}}\,.
\end{align}
Here, we find
\begin{align}
    g(a_i) = -a_i^2-A^2B^2/a_i^2\,,
\end{align}
with $a_0=A$ and $a_i=a_0+i$, $i=0,1,\dots,|N|$, which yields the energies of the partner Hamiltonians \cite{dabrowska1988explicit}
\begin{equation}
	\begin{split}
		&\varepsilon = A^2 + B^2 - (A-|N|/L)^2 - \frac{A^2 B^2}{(A-|N|/L)^2}\,.
	\end{split}
\end{equation}
Noting, in this case
\begin{align}\label{superpotentialen2}
 \varepsilon = E^2/\left[\hbar^2v_F^2(1-\kappa^2)\right],
\end{align}
we arrive at Eq.~\ref{energytanhelectric}.

\end{document}